\documentclass[twocolumn, showpacs, preprintnumbers,amsmath,amssymb]{revtex4}
\usepackage{graphicx}
\usepackage{color}

\newcommand{\grp}[1]{\mathrm{#1}}

\newcommand{\grU}{\grp{U}}

\newcommand{\abs}[1]{\lvert #1 \rvert}

\newcommand{\tr}{{\rm tr} }

\def\be{\begin{equation}}
\def\ee{\end{equation}}

\def\vol{{\mbox{vol}}}

\def\Uone{{\rm U}(1)}
\def\SUN{{\rm SU}(N)}

\begin{document}

\preprint{PUPT-2306}

\title{
Superconductors from Superstrings
}

\author{
Steven S.~Gubser, Christopher P.~Herzog, Silviu S.~Pufu, and Tiberiu Tesileanu
}

\date{
July 2009
}

\affiliation{
Department of Physics, Princeton University, Princeton, NJ 08544, USA
}

\begin{abstract}
We establish that in a large class of strongly coupled 3+1 dimensional ${\mathcal N}=1$ quiver conformal field theories with gravity duals, adding a chemical potential for the R-charge leads to the existence of superfluid states in which a chiral primary operator of the schematic form ${\cal O} = \lambda\lambda + {\mathcal W}$ condenses. 
Here $\lambda$ is a gluino and ${\mathcal W}$ is the superpotential. Our argument is based on the construction of a consistent truncation of type IIB supergravity that includes a $\Uone$ gauge field and a complex scalar.
\end{abstract}


\maketitle

{\it Introduction --}
Using the AdS/CFT correspondence, refs.\ \cite{Gubser:2008px,Hartnoll:2008vx} argued that a classical scalar-gravity model 
describes a superconducting phase transition in a dual strongly interacting field theory.  
Superfluid phase transition is perhaps a more accurate description \cite{Herzog:2008he} as there is no Higgs mechanism in the field theory, but for many physical questions, the distinction is irrelevant \cite{Hartnoll:2008kx}.
The proposal is interesting because it 
suggests that string theory techniques provide good theoretical control over superfluid transitions in certain strongly coupled theories, raising the hope that one might extend lessons learned from such theories to real condensed matter systems.
In this letter, we embed the scalar-gravity model in type IIB string theory.  The embedding 
clarifies the microscopic nature of the 3+1 dimensional field theory dual.

The AdS/CFT correspondence provides a recipe for constructing a large class of ${\mathcal N}=1$ supersymmetric, 3+1 dimensional conformal field theories (SCFTs) by placing a stack of $N$ D3-branes at the tip of a three complex dimensional Calabi-Yau cone $X$ in type IIB string theory
\cite{Kehagias:1998gn, Klebanov:1998hh, Acharya:1998db, Morrison:1998cs}. 
The field theory can be thought of as the open 
string degrees of freedom propagating on the D-branes at the Calabi-Yau singularity, 
and is a quiver gauge theory with $\SUN$ gauge groups and superpotential ${\cal W}$.

The AdS/CFT correspondence provides a dual closed string description of the field theory as type IIB string theory in the background curved by the energy density of the stack of D3-branes.  In the near horizon limit, i.e.\ close to the D3-branes, the ten dimensional space factorizes into a product of anti-de Sitter space and a Sasaki-Einstein manifold, $\text{AdS}_5 \times Y$, where $Y$ is a level surface of the cone $X$.  The R-symmetry of the SCFT is realized geometrically as an isometry of $Y$.

In section \ref{sec:consistent}, given a Sasaki-Einstein manifold $Y$ expressible as a $\grp U(1)$ fibration over a compact K\"ahler-Einstein base,
we write down a consistent 
truncation of type IIB supergravity to five dimensions 
that includes a complex scalar field $\Psi$
and the gauge field $A_\mu$ 
dual to the R-symmetry current.  
The field $\Psi$ is dual to a chiral primary operator ${\cal O}$ 
with scaling dimension $\Delta=3$ in the field theory.
In the presence of a chemical potential $\mu$, realized geometrically as the boundary value of 
$A_t$, the chiral primary develops an expectation value below a critical temperature 
$T_0$.  In the dual gravity language, an electrically charged black hole develops scalar hair.  By calculating the free energy as a function of $T$, we demonstrate that this phase transition that spontaneously breaks the $\Uone$ R-symmetry is second order.  

In section \ref{sec:general}, we calculate the critical temperature $T_p$ below which a more general complex scalar with conformal dimension $\Delta$ and R-charge $R$ will develop a perturbative instability.  In some cases, for example for 
$\Psi$ of section \ref{sec:consistent}, this $T_p$ corresponds to the critical temperature of a second order phase transition.  However, if the phase transition is first order, $T_c > T_p$ and $T_p$ is instead the temperature below which the symmetry restored phase of the field theory becomes perturbatively unstable.  We show that of all scalar chiral primary operators, ${\cal O}$ has the largest $T_p$ if it has lowest conformal dimension. 
If the latter condition is satisfied, 
then for reasons presented in section~\ref{sec:universal}, 
it is likely that
the condensation of ${\cal O}$ is responsible for a $\Uone$ symmetry breaking phase transition in the field theory.  
${\cal O}$ is at least tied for lowest conformal dimension in some quiver theories: in section~\ref{sec:universal} we give a particular example based on a ${\mathbb Z}_7$ orbifold of $S^5$. 

Embeddings in M-theory of 2+1-dimensional versions of these field theory models have been discussed in \cite{Denef:2009tp}.  While ref.\ \cite{Denef:2009tp} treats explicitly a broader range of examples than we do, their analysis of the scalar instability is perturbative.  Our consistent truncation allows us to establish the phase transition is second order and to follow the broken phase to arbitrarily low temperatures. Because we are working with a 3+1 dimensional field theory where the AdS/CFT duality is better understood, we are able to say more about the microscopic nature of the field theory dual than has been possible thus far in the 2+1 dimensional case. 

\subsection{A Consistent Truncation}
\label{sec:consistent}

Consider the following five-dimensional action, involving a metric, a $\Uone$ gauge field, and a complex scalar:
\be
S = \frac{1}{2 \kappa_5^2} \int d^5x \sqrt{-g} \left( {\mathcal L}_{\rm EM} + {\mathcal L}_{\rm scalar}\right) \,,
\ee
where
\be
\label{5dEM}
{\mathcal L}_{\rm EM} =  R  -
 \frac{L^2}{3}  F_{\mu\nu} F^{\mu\nu} 
+\left(\frac{2 L}{3}\right)^3 \frac{1}{4} \epsilon^{\lambda \mu\nu\sigma \rho} F_{\lambda \mu} F_{\nu \sigma} A_\rho  \,,
\ee
and \footnote{%
 S.~Franco, A.~Garcia-Garcia and D.~Rodriguez-Gomez,
  ``A general class of holographic superconductors,''
  arXiv:0906.1214 [hep-th] 
 considered a general class of scalar kinetic terms to which this example belongs.
}
\begin{eqnarray}
\label{5dscalar}
{\mathcal L}_{\rm scalar} = - \frac{1}{2} \Bigl[
(\partial_\mu \eta)^2 + \sinh^2 \eta ( \partial_\mu \theta - 2  A_\mu)^2  \nonumber \\
- \frac{6}{L^2} \cosh^2 \frac{\eta}{2} \left( 5 - \cosh \eta \right) \Bigr] \,.
\end{eqnarray}
We define $\epsilon^{01234} \equiv 1/\sqrt{-g}$.
The real fields $\eta$ and $\theta$ are the modulus and phase of the complex scalar $\Psi$.
Note that for small $\eta$, the potential term expands to yield
\be
V(\eta) = - \frac{12}{L^2} - \frac{3}{2L^2} \eta^2 + \ldots \,.
\ee
The leading order term comes from a negative cosmological constant, $\Lambda = - 6/L^2$.  
The second order term is a mass for the scalar, $m^2 L^2 = \Delta (\Delta-4) = -3$, and 
so $\Delta = 3$.
The $\Uone$ gauge field 
has been normalized such that ${\cal W}$ has R-charge 2 and chiral primary operators satisfy the relation $\Delta = 3 |R|/2$
\cite{Berenstein:2002ke}.  Our operator ${\cal O}$ has R-charge $R=2$ 
and is indeed chiral primary.

We claim that any solution to the classical equations of motion following from this action lifts to a solution of type IIB supergravity.
The lift resembles the consistent truncations of ref.\ \cite{Gauntlett:2009zw}, and it also generalizes the Pope-Warner type compactifications of 
type IIB SUGRA \cite{Romans:1984an}.
The ten dimensional metric for the lift has the form
\be
\label{10dMetric}
ds^2 = \cosh \frac{\eta}{2} \, ds_M^2 + \frac{L^2}{\cosh \frac{\eta}{2}} \left[ ds_V^2 + \cosh^2 \frac{\eta} {2} (\zeta^A)^2 \right] \,.
\ee
The metric on the manifold $M$ is a solution to the five dimensional equations of motion, while $V$ is a two complex dimensional manifold with a K\"ahler-Einstein metric $g_{a \bar b}$ such that $R_{a \bar b} = 6 g_{a \bar b}$.  Let $\omega$ be the K\"ahler form on $V$.  We construct a $\Uone$ fiber bundle over $V$ with the one form $\zeta^A = \zeta + 2A/3$ and $\zeta = d \psi + \sigma$ such that $d\zeta = 2 \omega$.

In the case $A=\eta=0$, the line element $ds_Y^2 = ds_V^2 + \zeta^2$ on the five dimensional space $Y$ is a Sasaki-Einstein metric.  Moreover, introducing a radial coordinate $r>0$, 
the line element $dr^2 + r^2 ds_Y^2$
is a Ricci flat metric on a K\"ahler manifold $X$ with a conformal scaling symmetry $r \to \lambda r$.  In other words, $X$ is a Calabi-Yau cone.

Denoting $F = dA$ and $J =  \sinh^2 \eta (d\theta - 2 A)$, we can write the self-dual five-form as
 \be
 \label{F5SE}
  F_5 = \frac{1}{g_s} \left( {\cal F} + *{\cal F} \right) \,,
 \ee
where
 \begin{eqnarray}
 \label{GotFCalFAgain}
   {\cal F} &\equiv& -\frac{1}{L} \cosh^2 {\eta \over 2} \left(\cosh \eta - 5\right) {\rm vol}_M
    -\frac{2 L^3}{3} (*_M F) \wedge \omega \nonumber \\
    &&
    + {L^4 \over 4  \cosh^4 {\eta \over 2}} J \wedge \omega^2 \  , \\
  *{\cal F} &=& L^4 { \left(\cosh \eta - 5\right) \over 2  \cosh^2 {\eta \over 2}}
    \zeta^A \wedge \omega^2 
    + \frac{2 L^4}{3} F \wedge \zeta^A \wedge \omega \nonumber \\
    &&
    + {L \over 2} (*_M J) \wedge \zeta^A \,. 
 \end{eqnarray}
To specify the two-form gauge potentials we first consider the holomorphic $3$-form $\hat \Omega_3$ on the Calabi-Yau three-fold $X$ normalized so that ${\hat \Omega}_3 \wedge {\hat {\bar \Omega}}_3 = 8 \vol_X$.  This form decomposes as
 \be
 \label{hatOmega3DecompSE}
  \hat \Omega_3 = r^3 \left({dr \over r} \wedge \Omega_2 + \Omega_3 \right) \,,
 \ee
where $r$ is the radial coordinate of the cone.  The form $e^{-3 i \psi} \Omega_2$ is a primitive $(2, 0)$ form on $V$ satisfying $d(e^{-3 i \psi} \Omega_2) = 3 i e^{-3 i \psi} \sigma \wedge \Omega_2$ and $\Omega_2 \wedge \bar \Omega_2 = 2 \omega^2$ \cite{Martelli:2004wu}. 
The two-form potentials are
 \be
 \label{F2SE}
  F_2 = {L^2} \tanh \frac{\eta}{2} e^{i \theta} \Omega_2 \,,
    \qquad F_2 \equiv  B_2 + i g_s C_2 \,.
 \ee

One can check that the ansatz given by \eqref{10dMetric}, \eqref{F5SE}, and \eqref{F2SE} leads to a consistent reduction with the effective five-dimensional lagrangian given by \eqref{5dEM} and \eqref{5dscalar}.

\vskip0.1in

{\it The Phase Transition --}
We are interested in studying the response of an SCFT to an R-charge chemical potential $\mu$ and a temperature $T$.  
One expects that for low enough $T/\mu$, R-charged operators develop expectation values
that spontaneously break the $\grU(1)$ R-symmetry. 
At high temperatures, the field theory is dual to an electrically charged black hole in anti-de Sitter space, while at a sufficiently low temperature, the black-hole acquires scalar hair \cite{Gubser:2008px,Hartnoll:2008vx}.  The electrically charged black hole which solves the equations of motion following from (\ref{5dEM}), 
along with a negative cosmological constant $\Lambda = -6/L^2$, takes the form
\be
 \begin{split}
   ds^2 &= \frac{L^2}{z^2} \left[ -f(z) \, dt^2 + d \vec x^2 + \frac{dz^2}{f(z)} \right] \,, \\
   A &= \mu \biggl[1 - \biggl(\frac{z}{z_h} \biggr)^2 \biggr] dt \,,
 \end{split}
\label{RNbg}
\ee
where
\be
f(z) = 1 + Q^2 \biggl( \frac{z}{z_h} \biggr)^6 - (1+Q^2) \biggl( \frac{z}{z_h} \biggr)^4 \,, 
\ee
with the charge $Q= 2 z_h \mu/3$.  The Hawking temperature of the black hole is
$
T_H = (2-Q^2)/2 \pi z_h 
$.

\begin{figure}[h]
\begin{center}
 \includegraphics[width=3.5in]{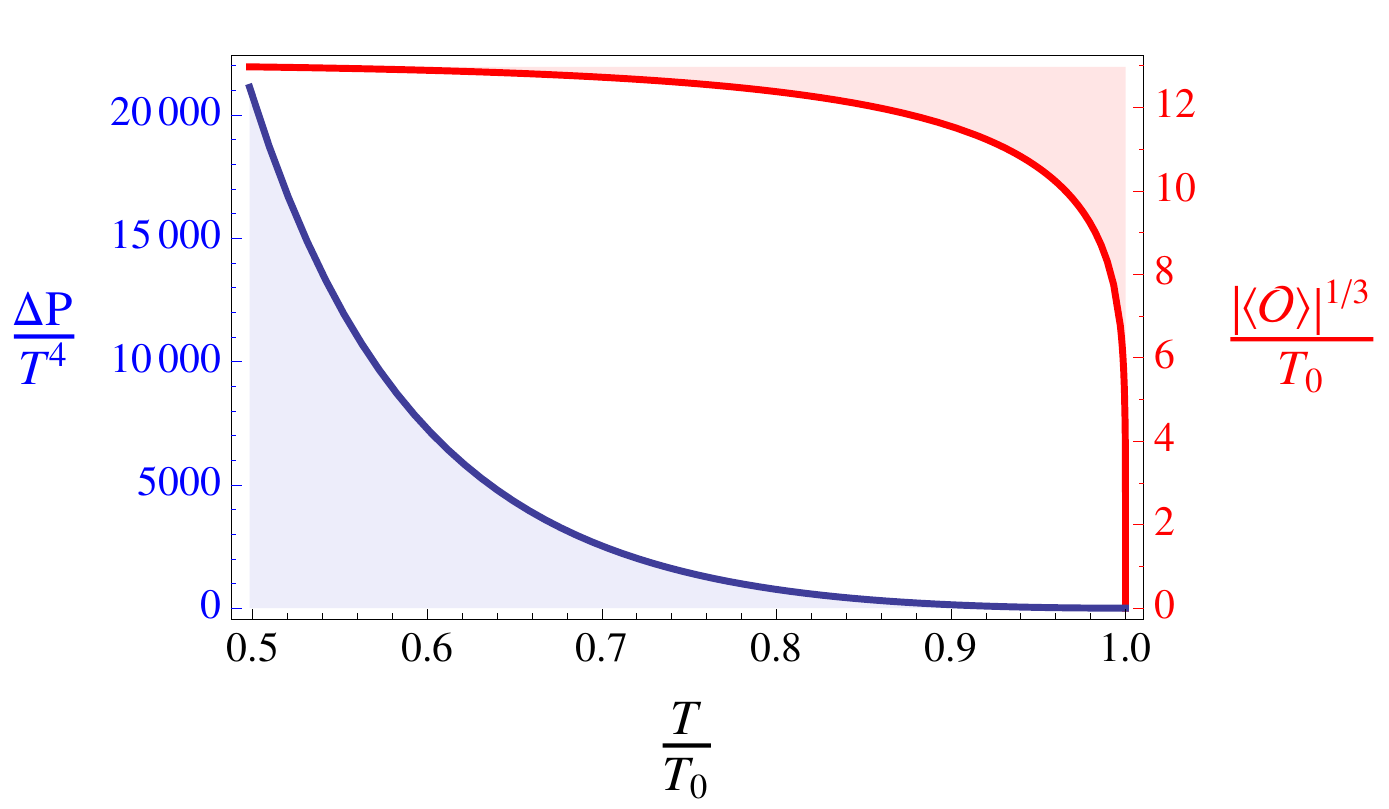}
\end{center}
\caption{
(Color online.)  Upper right plot: $\abs{\langle {\cal O} \rangle}^{1/3} / T_0$ vs.\ $T/T_0$, where $\langle {\cal O} \rangle$ is expressed as multiples of $L^3/ \kappa_5^2$.  The critical temperature is $T_0 \approx 0.0607 \mu$.  Near $T_0$, $\langle {\cal O} \rangle \sim \abs{T - T_0}^{1/2}$, indicating a mean field critical exponent.  Lower left plot: $\Delta P / T^4$ vs.\ $T/T_0$, where $\Delta P$ is the difference in pressure between the broken and unbroken phases, calculated in the grand canonical ensemble.  Near $T_0$, one has $\Delta P \sim (T-T_0)^2$, so the phase transition is second order.}\label{fig:BreakingPlot}
\end{figure}

At low temperatures, 
a hairy black hole solution with $\eta \neq 0$. 
becomes available.  
We find this solution numerically, 
using the techniques described, e.g., in \cite{Hartnoll:2008kx}.  
By a gauge choice, we can set $\theta =0$.  
We require no deformation of the conformal field theory by the symmetry breaking operator ${\mathcal O}$ dual to $\eta$.  So for small $z$, 
$\eta = 
- z^3 ( \langle {\cal O} \rangle  \kappa_5^2 /L^3+ \ldots) $,
and the expectation value $\langle {\cal O} \rangle$ 
is the order parameter for breaking the $U(1)$ symmetry. 
With a black hole horizon at $z=z_h$, 
the other boundary conditions we impose are that $A_t(z{=}0) = \mu$ and $A_t(z{=}z_h) = 0$.
Fig.\ \ref{fig:BreakingPlot} gives $\langle {\cal O} \rangle$ as a function of $T$. 

We also plot the difference in pressure between the electrically charged black hole phase and the hairy black hole phase in fig.~\ref{fig:BreakingPlot}.  The pressure is a coefficient in the near boundary expansion of $g_{tt}$:
\be
g_{tt} = -\frac{L^2}{z^2} +  2 \kappa_5^2 P \frac{z^2}{L} + \ldots \,.
\ee
It is related to the free energy via $\Omega = - P\, \mbox{Vol}$.
To within our numerical precision, $\partial \Delta \Omega / \partial T=0$ at $T=T_0$, indicating a second order phase transition.

We have also calculated the conductivity for this model as a function of frequency and $T$, 
using the techniques of ref.\
\cite{Hartnoll:2008vx}.  The results are qualitatively similar to those of a free $\Delta =3$ scalar in the probe limit \cite{Horowitz:2008bn}.

\subsection{Perturbative Instabilities}
\label{sec:general}

\begin{figure}[h]
\begin{center}
 \includegraphics[width=3in]{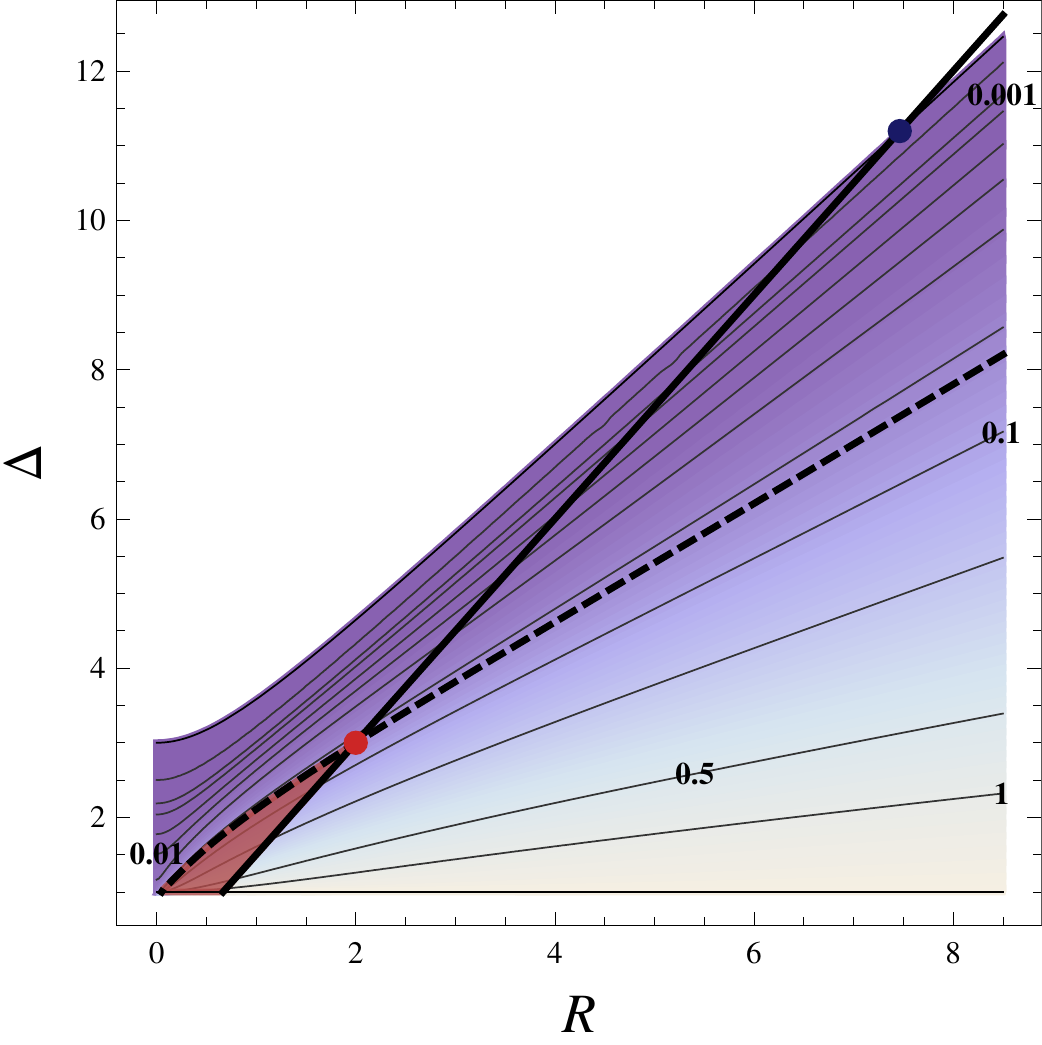}
\end{center}
\caption{(Color online.)  A contour plot of $T_p/\mu$ as a function of $\Delta$ and $R$.  The numbers next to the contour lines represent $T_p/\mu$.
We need only consider scalars above the unitarity bound, $\Delta \geq 1$ \cite{Mack:1975je}.
The dark solid line is the BPS bound $\Delta = 3 R/2$ \cite{Dobrev:1985qv}.
Scalars which are 
less stable than the operator $\mathcal O$ 
are restricted to the triangular, shaded region near the lower-left corner.
\label{fig:condensate}}
\end{figure}

Although our chiral primary leads to spontaneous breaking of the U(1) R-symmetry at low temperature if $\mu \neq 0$, these SCFT's typically have many operators with R-charge.  It may be that there exists another operator which produces a phase transition at a higher $T$.  
Such an operator need not be a scalar.  We focus on the case that this ``less stable'' operator is a scalar, and 
replace ${\mathcal L}_{\rm scalar}$ with 
\be
- \frac{1}{2}\left[ (\partial_\mu \eta)^2 + \eta^2 ( \partial_\mu \theta - R A_\mu)^2  + m^2 \eta^2\right] + \frac{12}{L^2}\,,
\ee
representing the leading quadratic terms for a complex scalar of charge $R$ and conformal dimension satisfying $m^2 L^2 = \Delta (\Delta -4)$.  

These leading quadratic terms are enough to calculate the temperature $T_p$ below which the electrically charged black hole becomes perturbatively unstable with respect to an exponentially growing mode of the scalar.  We calculate $T_p$ by looking for a zero mode solution of $\eta$ in the electrically charged black hole background (\ref{RNbg}).  Such a solution will have the leading behavior $\eta \sim z^\Delta$ near the boundary and should be finite at the horizon.  
 Depending on higher order terms in the scalar action, which in general we don't know, $T_p$ may be the point of an actual second order phase transition, as it was in section \ref{sec:consistent}, or it may label a spinodal point in a first order phase transition, beyond which the electrically charged black hole ceases to be perturbatively stable.   

We solved for this zero mode numerically, and the resulting $T_p$ is plotted in 
fig.~\ref{fig:condensate} as a function of $\Delta$ and $R$.  
As described in \cite{Denef:2009tp}, there is a critical curve in the $R$-$\Delta$ plane where $T_p=0$.  This curve can be determined analytically by considering the behavior of the scalar field in an extremal electrically charged black hole solution:
\be
R^2 = \frac{2}{3} \Delta(\Delta-4) +2 \,.
\ee 
Note the BPS line $\Delta = 3 R/2$ intersects
this curve,
leaving a region of finite area in the $R$-$\Delta$ plane with $T_p>0$.

Note that $T_p$ is a monotonically increasing function of $R$ and a monotonically 
decreasing function of $\Delta$.  Moreover, along the BPS bound $\Delta = 3 R / 2$,
$T_p$ is a decreasing function of $\Delta$.  These results suggest that the superfluid phase transition will be caused by an operator at or close to the BPS bound and of small $\Delta$.

We found in section \ref{sec:consistent} that for SCFT's dual to D3-branes at Calabi-Yau singularities, 
there always exists an operator
with $\Delta =3$ that saturates the BPS bound.  Given the existence of such an operator and corresponding $T_0$, scalars which are perturbatively less stable, 
should they exist, are restricted to a tiny corner of the
$R$-$\Delta$ plane; see fig.~\ref {fig:condensate}.
We can in general check if there are any chiral primary operators with $\Delta<3$.  It is less straightforward to rule out unprotected operators with a $T_c >T_0$.  Nevertheless, the expectation is that such operators do not exist, at least for strongly interacting theories with AdS/CFT duals.   
Large couplings are associated with large anomalous dimensions.  In these AdS/CFT constructions, we expect generic operators to be dual to string states with masses of order the string scale, $m \ell_s \sim 1$, and thus $\Delta \sim (g_s N)^{1/4}$.   It seems unlikely we will find any unprotected operator with a $T_c > T_0$.

\subsection{A Universal Chiral Primary Operator}
\label{sec:universal}

What chiral primary operators in an SCFT have $R=2$ and $\Delta =3$?  
Our supergravity solutions are dual to superconformal quiver gauge theories via the AdS/CFT correspondence.  
The lowest component ${\mathcal O}$ of the F-term 
in the Lagrangian describing these field theories takes the form
 \be\label{WhatIsO}
  {\mathcal O} = {\mathcal W}(\phi_i) + 
    {1 \over 32\pi i} \sum_j \tau_j \, \tr \,  \lambda_{j\alpha}^2 \,,
 \ee
where ${\mathcal W}$ is the superpotential, 
the gluino field $\lambda_{\alpha}$ is the lowest component of the superfield
$W_\alpha$, and
the complex scalar fields $\phi_i$ 
are the lowest component of the chiral matter superfields $\Phi_i$.  The $\tau_j = \theta_j/ 2\pi + 4 \pi i / g_j^2$ are the complexified gauge couplings, and the sum $j$ runs over the gauge groups in the quiver.  
Because of conformal invariance and holomorphy arguments, we expect ${\mathcal O}$ to be a protected operator (up to non-perturbative corrections, wave-function rescaling, and possible mixing with a descendant of the Konishi operator).  
It is true that ${\mathcal F}$ has $R=2$ and $\Delta =3$.  We claim that ${\mathcal O}$ is dual to the complexified scalar supergravity field $(\eta, \theta)$. 



Consider the cases where the Sasaki-Einstein manifold is the sphere $Y=S^5$ or the level surface of the conifold $Y = T^{1,1}$.  In the first case, the dual field theory is ${\mathcal N}=4$ 
$\SUN$ SYM.  In ${\mathcal N}=1$ notation, the field theory has three chiral superfields $X$, $Y$, and $Z$ transforming in the adjoint representation of $\SUN$, and a superpotential
${\mathcal W} \sim \tr \, X [Y,Z]$. 
In the second case, the $\SUN \times \SUN$ field theory has  bifundamental fields $A_i$ and $B_i$, $i=1,2$ transforming under the $(N, \bar N$) and $(\bar N, N)$ representations of the gauge groups and a superpotential ${\mathcal W} \sim \epsilon_{ij} \epsilon_{kl} \tr (A_i B_k A_j B_l)$.  In both these cases, ${\mathcal O}$ is indeed dual to the complexified scalar $(\eta, \theta)$ \cite{Ceresole:1999zs}. 

Note that for $S^5$ and $T^{1,1}$, 
${\mathcal O}$ will not cause the phase transition that breaks a U(1) R-symmetry.  The reason is that there exist chiral primary operators for these SCFT's with lower conformal dimension.  For $S^5$, $\tr (X^2)$ has $\Delta = 2$
while for $T^{1,1}$, $\tr (A_i B_j)$ has $\Delta = 3/2$, and both of these operators
condense at a higher $T_c$.  Thus, we need to look for SCFTs where ${\cal O}$ has the lowest conformal dimension among the chiral primaries.

One such theory
is $S^5 / {\mathbb Z}_7$ where the orbifold acts with weights $(1,2,4)$ on the 
${\mathbb C}^3 \supset S^5$. 
The quiver field theory has $G= \SUN^7$.  The three chiral superfields $X$, $Y$, and $Z$ of ${\mathcal N}=4$ SYM become 21 fields $X^i_{i+1}$, $Y^i_{i+2}$, and $Z^i_{i+4}$.  Here $X^{i}_{j}$ indicates a field that transforms under the anti-fundamental of the $i^{\rm th}$ gauge group and the fundamental of the $j^{\rm th}$ and $X^{i+7}_{j} = X^{i}_{j}$.  
A chiral primary tied for smallest conformal dimension in this field theory is ${\mathcal O}$.  (The other chiral primary is related to the beta deformation \cite{Lunin:2005jy}.) 
More generally, we expect an orbifold $S^5 / {\mathbb Z}_n$ with weights $(w_1, w_2, w_3)$ such that $n=w_1+w_2+w_3$ to be a candidate provided that the $w_i$ are distinct and that $w_i \neq -w_j \mod n$ for all $i$ and $j$.

\vskip 0.1in

{\it Acknowledgments --}
We would like to thank D.~Berenstein, F.~Denef, T.~Dumitrescu, S.~Hartnoll, I.~Klebanov, F.~Rocha, N.~Seiberg, and A.~Yarom for discussion.  
This work was supported in part by the US NSF under Grant Nos.\ PHY-0551164 (CH), -0652782 (SG, SP), -0756966 (CH), and -0844827 (CH, SP, TT) and also by the DOE under Grant No.\ DE-FG02-91ER40671 (SG).


\begin{thebibliography}{99}

\bibitem{Gubser:2008px}
  S.~S.~Gubser,
  Phys.\ Rev.\  D {\bf 78}, 065034 (2008)
  [arXiv:0801.2977 [hep-th]].

\bibitem{Hartnoll:2008vx}
  S.~A.~Hartnoll, C.~P.~Herzog and G.~T.~Horowitz,
  Phys.\ Rev.\ Lett.\  {\bf 101}, 031601 (2008)
  [arXiv:0803.3295 [hep-th]].
  
  \bibitem{Herzog:2008he}
  C.~P.~Herzog, P.~K.~Kovtun and D.~T.~Son,
  Phys.\ Rev.\ D {\bf 79}, 066002 (2009)
  [arXiv:0809.4870 [hep-th]].

\bibitem{Hartnoll:2008kx}
  S.~A.~Hartnoll, C.~P.~Herzog and G.~T.~Horowitz,
  JHEP {\bf 0812}, 015 (2008)
  [arXiv:0810.1563 [hep-th]].

  
  \bibitem{Kehagias:1998gn}
  A.~Kehagias,
  Phys.\ Lett.\  B {\bf 435}, 337 (1998)
  [arXiv:hep-th/9805131].

\bibitem{Klebanov:1998hh}
  I.~R.~Klebanov and E.~Witten,
  Nucl.\ Phys.\  B {\bf 536}, 199 (1998)
  [arXiv:hep-th/9807080].

\bibitem{Acharya:1998db}
  B.~S.~Acharya, J.~M.~Figueroa-O'Farrill, C.~M.~Hull and B.~J.~Spence,
  Adv.\ Theor.\ Math.\ Phys.\  {\bf 2}, 1249 (1999)
  [arXiv:hep-th/9808014].

\bibitem{Morrison:1998cs}
  D.~R.~Morrison and M.~R.~Plesser,
  Adv.\ Theor.\ Math.\ Phys.\  {\bf 3}, 1 (1999)
  [arXiv:hep-th/9810201].

 \bibitem{Denef:2009tp}
  F.~Denef and S.~A.~Hartnoll,
  ``Landscape of superconducting membranes,''
  arXiv:0901.1160 [hep-th].

 \bibitem{Berenstein:2002ke}
  D.~Berenstein, C.~P.~Herzog and I.~R.~Klebanov,
  JHEP {\bf 0206}, 047 (2002)
  [arXiv:hep-th/0202150].

\bibitem{Gauntlett:2009zw}
  J.~P.~Gauntlett, S.~Kim, O.~Varela and D.~Waldram,
  JHEP {\bf 0904}, 102 (2009)
  [arXiv:0901.0676 [hep-th]].

\bibitem{Romans:1984an}
 L.~J.~Romans,
 Phys.\ Lett.\  B {\bf 153}, 392 (1985).

\bibitem{Martelli:2004wu}
  D.~Martelli and J.~Sparks,
  Commun.\ Math.\ Phys.\  {\bf 262}, 51 (2006)
  [arXiv:hep-th/0411238].


\bibitem{Mack:1975je}
  G.~Mack,
  Commun.\ Math.\ Phys.\  {\bf 55}, 1 (1977).
  
  \bibitem{Dobrev:1985qv}
  V.~K.~Dobrev and V.~B.~Petkova,
  Phys.\ Lett.\  B {\bf 162}, 127 (1985).

\bibitem{Horowitz:2008bn}
  G.~T.~Horowitz and M.~M.~Roberts,
  Phys.\ Rev.\  D {\bf 78}, 126008 (2008)
  [arXiv:0810.1077 [hep-th]].

\bibitem{Ceresole:1999zs}
  A.~Ceresole, G.~Dall'Agata, R.~D'Auria and S.~Ferrara,
  Phys.\ Rev.\  D {\bf 61}, 066001 (2000)
  [arXiv:hep-th/9905226].
  
  \bibitem{Lunin:2005jy}
  O.~Lunin and J.~M.~Maldacena,
  JHEP {\bf 0505}, 033 (2005)
  [arXiv:hep-th/0502086].

\end{thebibliography}
\end{document}